\documentclass[sigconf]{acmart}

\AtBeginDocument{%
  \providecommand\BibTeX{{%
    \normalfont B\kern-0.5em{\scshape i\kern-0.25em b}\kern-0.8em\TeX}}}

\copyrightyear{2022}
\acmYear{2022}
\setcopyright{acmcopyright}
\acmConference[FAccT '22]{2022 ACM Conference on Fairness, Accountability, and Transparency}{June 21--24, 2022}{Seoul, Republic of Korea}
\acmBooktitle{2022 ACM Conference on Fairness, Accountability, and Transparency (FAccT '22), June 21--24, 2022, Seoul, Republic of Korea}
\acmPrice{15.00}
\acmDOI{10.1145/3531146.3533219}
\acmISBN{978-1-4503-9352-2/22/06}

\usepackage{makecell}
\usepackage{booktabs}

\usepackage{colortbl}

\usepackage{tikz}
\usetikzlibrary{arrows}

\begin{document}

\title{Characterizing Properties and Trade-offs of Centralized Delegation Mechanisms in Liquid Democracy}

\author{Brian Brubach}
\email{bb100@wellesley.edu}
\affiliation{%
  \institution{Wellesley College}
  \city{Wellesley}
  \state{MA}
  \country{USA}
}

\author{Audrey Ballarin}
\email{ab14@wellesley.edu}
\affiliation{
  \institution{Wellesley College}
  \city{Wellesley}
  \state{MA}
  \country{USA}
}

\author{Heeba Nazeer}
\email{hn1@wellesley.edu}
\affiliation{
  \institution{Wellesley College}
  \city{Wellesley}
  \state{MA}
  \country{USA}
}







\renewcommand{\shortauthors}{Brubach, et al.}

\begin{abstract}
    Liquid democracy is a form of transitive delegative democracy that has received a flurry of scholarly attention from the computer science community in recent years. In its simplest form, every agent starts with one vote and may have other votes assigned to them via delegation from other agents. They can choose to delegate all votes assigned to them to another agent or vote directly with all votes assigned to them. However, many proposed realizations of liquid democracy allow for agents to express their delegation/voting preferences in more complex ways (e.g., a ranked list of potential delegates) and employ a centralized delegation mechanism to compute the final vote tally. In doing so, centralized delegation mechanisms can make decisions that affect the outcome of a vote and where/whether agents are able to delegate their votes. Much of the analysis thus far has focused on the ability of these mechanisms to make a correct choice. We extend this analysis by introducing and formalizing other important properties of a centralized delegation mechanism in liquid democracy with respect to crucial features such as accountability, transparency, explainability, fairness, and user agency. In addition, we evaluate existing methods in terms of these properties, show how some prior work can be augmented to achieve desirable properties, prove impossibility results for achieving certain sets of properties simultaneously, and highlight directions for future work.
\end{abstract}

\ccsdesc[500]{Applied computing~Voting / election technologies}
\ccsdesc[500]{Applied computing~E-government}
\ccsdesc[300]{Theory of computation~Algorithmic game theory and mechanism design}

\keywords{liquid democracy, voting, computational social choice, fairness, accountability, transparency}


\maketitle

\section{Introduction}

Originally championed by Charles Dodgson (AKA Lewis Carroll) in 1884~\citep{carroll1884principles,ramos2015liquid}, liquid democracy is an old idea that has attracted renewed interest from both computer science~\citep{kahng2021algorithmic,brill2018interactive,golz2018fluid,brill2018pairwise,caragiannis2019contribution,zhang2021tracking,kotsialou2018incentivising} and political philosophy~\citep{blum2016liquid} perspectives.  The basic concept is simple: it is a voting system in which each agent can choose to either vote directly or delegate their vote(s) to another agent. However, unlike the more common and restricted system of proxy voting, delegation in liquid democracy is transitive: If some votes are delegated to an agent, that agent may choose to delegate all of those votes (and their own vote) to another agent.

As liquid democracy has become a topic of serious study in the computer science and artificial intelligence communities, works have studied its ability to make a correct choice~\citep{kahng2021algorithmic,caragiannis2019contribution} as well as how best to collect and aggregate agents' delegation preferences~\citep{golz2018fluid,colley2021smart,zhang2021tracking}. It was implemented internally at Google~\citep{hardt2015google}, by Germany's Pirate Party~\citep{kling2015liquidfeedback}, and in several other cases~\citep{paulin2020overview}. The mechanisms proposed and studied in these works allow agents to specify delegation preferences in a variety of ways (e.g., providing a ranked or unranked list of acceptable delegates) and sometimes include additional objectives/constraints such as limiting the total number of votes delegated to any single individual. Although still far from widespread adoption, these real-world applications and increased study illustrate the promise of liquid democracy and motivate deeper examination.

The essential idea of liquid democracy may be straightforward to state, but there are many ways to translate it into practice. The seminal work of \citet{blum2016liquid} listed four key components:
\begin{enumerate}
    \item \emph{Direct democracy:} Every agent has the option to vote directly on an issue.
    \item \emph{Flexible delegation:} Agents can delegate their votes to others for a single policy issue, all issues in a policy area, or all issues.
    \item \emph{Meta-delegation:} Agents can also delegate any votes that have been delegated to them (delegation is transitive).
    \item \emph{Instant recall:} An agent can withdraw a delegation at any time.
\end{enumerate}
Those four components suffice to establish the basic model of liquid democracy. However, they also leave several ambiguities which should be resolved and made explicit in any transparent implementation. We refer the reader to \citet{blum2016liquid} for a discussion of many such challenges that can arise from a political science perspective as well as which applications are best-suited to this form of governance and why.

Our work seeks to bring a fairness, accountability, and transparency lens to the computational study of mechanisms for translating agent preferences into outcomes within realizations of liquid democracy. 
Now that there are many proposed (and in some cases implemented) mechanisms for liquid democracy systems, how do we choose between them? Individual mechanisms have been analyzed in terms of certain key features and the qualities they were designed to support, but not thoroughly compared among a broad range of other important criteria. In particular, many questions arise which are not adequately addressed in prior work such as,
\begin{enumerate}
    \item What rights should agents have? Should they have the right to delegate instead of voting directly, and if so, can the mechanism guarantee that? (Section~\ref{sec:righttodelegate}) 
    \item Can the mechanism itself arbitrarily choose the outcome of the vote and what governs its decision? (Section~\ref{sec:arbitrary}) 
    \item How transparent and interpretable is the mechanism's explanation of what happened to an agent's vote in a previous election, and how easily can agents predict the effect of changing their delegation preferences? (Sections~\ref{sec:posthocexplain} and~\ref{sec:localpredictable}) 
\end{enumerate}
We formalize these types of questions by characterizing several properties of liquid democracy mechanisms and exploring how these properties interact with each other.

\section{Outline}

Section~\ref{sec:prelim} provides some basic definitions and discusses related work not covered elsewhere in the paper. Section~\ref{sec:properties} lists and defines several properties which we use to analyze and compare mechanisms for liquid democracy. Section~\ref{sec:analysis} analyzes five prominent existing mechanisms in terms of the properties we propose. Finally, Section~\ref{sec:conclusion} provides a conclusion and future directions.

\section{Preliminaries and related work}
\label{sec:prelim}

\subsection{Basic definitions and notation}

In this section, we provide some basic terminology that will be used throughout the paper. We refer to all participants in a liquid democracy system as \emph{agents}. Any agent who prefers to cast a direct vote as opposed to delegating or abstaining is called a \emph{voter} and we limit our scope to mechanisms which always allow any such agent to cast a direct vote in keeping with that basic tenet of liquid democracy. We largely ignore the agents who abstain from both voting and delegation except when considering a switch from abstaining to delegating or voting over the course of multiple elections. We additionally limit the discussion in this paper to simple plurality votes on binary issues.

While we consider multiple ways in which agents can express their delegation and voting preferences, agent preference in this work are represented by some form of directed graph which we call a \emph{preference graph}. A directed edge from agent $u$ to agent $v$ in the preference graph means that $u$ is willing to delegate their vote to $v$ and prefers this delegation over voting directly. Edges in the preference graph may additionally be assigned ranks or weights as described in Section~\ref{sec:expression}. To determine how votes are actually delegated and cast, a \emph{tally algorithm} takes a preference graph as input and outputs a \emph{delegation graph} which explicitly describes how each agent's vote(s) are delegated or cast. This delegation graph determines how many will be assigned to each outcome.

\subsection{Modeling assumptions and limitations}
\label{sec:simpleassumptions}

Regarding agents decisions to vote directly, we make two minor simplifying assumptions. First, we assume that any agent who lists at least one approved delegate would prefer to delegate rather than vote directly. Second, while we explore the possibility that a mechanism cannot (or chooses not to) transfer an agent's vote(s) to any approved delegate, we do not discuss what to do with such agents. We view the following as the most natural options for handling this situation of agents whose votes are not delegated by the mechanism: request a direct vote from these agents, discard these agents, or allow all agents to specify a backup direct vote that can be applied in this case. Because these choices can be applied to any of the mechanisms we discuss, we do not see them as inherent properties of a given mechanism and assume that a mechanism designer will address this issue however they see fit. Our primary focus in this realm will instead be on whether an agent who prefers to delegate is allowed to delegate by the mechanism.

Another aspect we do not engage with is the ability of agents to specify preferences across multiple elections or issues at once. For example, LiquidFeedback~\citep{behrens2014principles} allows an agent to delegate to another agent globally (all issues), only for certain categories of issues, or only for specific issues. Google Votes~\citep{hardt2015google,hardt2014googletalk} similarly allows agents to tag potential delegates with categories. However, while these mechanisms collect such complex preferences for the sake of efficiency, they generally resolve preferences for a singe issue at a time using the type of preference graph we study (a simple, directed graph with ranked or unranked edges).

\subsection{Epistemic versus equality-based accounts}

\citet{blum2016liquid} suggests two distinct functions through which we can evaluate liquid democracy. The epistemic account assumes there is some ground truth correct choice and values the ability of a mechanism to find that correct choice. Equality-based accounts on the other hand emphasize the rights of all participants to have a voice and promote their own individual, subjective choice of preferred outcome. 

In some ways, these accounts complement each other. Agents want to select a delegate that represents their needs (equality-based) and makes competent decisions in support of those needs (epistemic). 
However, we will see that these values can also be somewhat competing. Epistemic approaches may favor mechanisms that exert heavier influence over agents in order to ideally increase the chance of making the ``correct'' choice. For example, \citet{golz2018fluid} suggests, ``If agents are considering multiple potential delegates, all of whom they trust, they are encouraged to leave the decision for one of them to a centralized mechanism.'' As such the relevance of our proposed properties can vary depending on which function one favors. For example, giving all agents a right to delegate may be valued more strongly from an equality-based perspective than an epistemic one. 
A further complication we explore is that mechanisms exerting a larger influence may obscure the information needed for agents to make intelligent delegation choices, thus potentially harming chances of correctness.

\subsection{Other related work}

Several prior works which propose centralized mechanisms are discussed at length in later sections. Due to our focus on simple binary votes, we do not analyze some proposals that interpret more complex preference expressions~\citep{colley2021smart} or ranked outcomes~\citep{brill2018pairwise}. Other works that give broad overviews of liquid democracy and compare approaches include \citet{brill2018interactive} and \citet{paulin2020overview} with the latter focusing especially on the real world implementations.

One prior work, analyzed in Section~\ref{sec:bfd}, investigates the concept of ``incentivising participation''~\citep{kotsialou2018incentivising} which has a similar flavor to some properties studied in the present work. In short, they argue that participation among voters (which they call \emph{guru participation}) is disincentivised when an agent can change their delegation preferences in such a way that the voter who ultimately casts that agent's delegated vote is unhappy with the change. However, the voter in this case can still be given the incentive to vote (which they call \emph{cast participation}) and cannot benefit directly from rejecting the vote delegated by the offending agent. Because the voter in this scenario still has an incentive to vote and we see the ability of agents to affect voter power as fundamental to liquid democracy, we do not include this property in our assessment. Further, we note that the voter who is considered ``disincentivised'' within this model may nevertheless prefer a mechanism which does not satisfy guru participation over a mechanism which does. Thus, while the observations in \citet{kotsialou2018incentivising} are surprising and interesting within the context of that work, it could be misleading for a practitioner to use this property in choosing between mechanisms.

\section{Properties of a Central Delegation Mechanism (CDM)}
\label{sec:properties}

This section outlines several important properties of a central delegation mechanism. Section~\ref{sec:expression} covers the most common ways that agents are able to express their preferences to the mechanism. 
Section~\ref{sec:influence} addresses key issues regarding how a mechanism can influence the outcome of a vote or limit an agent's ability to delegate. Section~\ref{sec:transexplainaccount} defines properties affecting the related goals of transparency, explainability, and accountability. Section~\ref{sec:runningtime} briefly discuss the effects of running time on the ability to satisfy other desirable properties. Finally, Section~\ref{sec:novelproperties} helps to clarify which properties are novel to this work.

\subsection{Expression of delegation preferences} 
\label{sec:expression}

One of the most obvious ways that mechanisms differ is in how they allow agents to express their delegation preferences. In our analysis, we will consider four common ways in which agents can specify their delegation preferences.
\begin{itemize}
    \item \emph{One neighbor preferences (ONP):} Agents may only choose one neighbor to delegate to. The preference graph is unweighted with out-degree $1$.
    \item \emph{Multiple ranked preferences (MRP):} Agents provide a ranked list of neighbors they are willing to delegate to. In the preference graph, the outgoing edges of each vertex are ranked.
    \item \emph{Multiple unranked preferences (MUP):} Agents provide an unranked list of neighbors they are willing to delegate to. The preference graph is unweighted.
\end{itemize}

In all cases, an agent who wishes to vote directly will have a single outgoing edge to one of the two binary choices. We note that the mechanism may bound the length of the list given by each agent in which case one neighbor preferences is a special case of both multiple ranked preferences and multiple unranked preferences with lists of length at most one.

As stated in Section~\ref{sec:simpleassumptions}, we will not cover how or whether a mechanism solicits votes from agents who prefer delegation over voting, but would like to vote if they can't delegate.

\subsection{Influence of the mechanism on outcomes and rights to delegation preferences}
\label{sec:influence}

Now, we introduce properties that address some of our central fairness questions: How and why can a mechanism influence the result of a vote and govern the agents' abilities to delegate? 
We begin by reviewing the common issue of delegation cycles and proposing definitions for an agent's right to delegate and to their top ranking. 
We then outline two key factors in mechanisms influencing an agent's right to delegate and/or the outcome of the vote: the goal of limiting concentrated power and the arbitrary decisions that can occur due to ambiguity in the mechanism objective.

\subsubsection{Resolving delegation cycles.} 
\label{sec:cycles}
\begin{figure}
\includegraphics{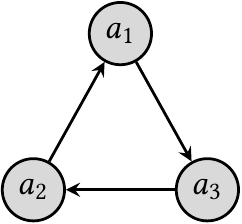}
    \caption{Example of a cycle in a preference graph. These agents' votes cannot be cast unless at least one of them becomes a voter.}
    \label{fig:triangle}
    \Description{A simple directed cycle on three nodes, a1, a2, and a3.}
\end{figure}

Delegation cycles are perhaps the most well-known challenge that arises when attempting to implement liquid democracy. As shown in Figure~\ref{fig:triangle}, we may receive delegation preferences that form a directed cycle (AKA a \emph{delegation cycle}). In this case, the mechanism must decide which agents (if any) are permitted to delegate. We note that prior work primarily addresses the issue of delegation cycles in one of the following three ways:
\begin{itemize}
    \item \emph{Assume away cycles (AAC):} In this case, the mechanism is designed for a model that assumes delegation cycles do not occur. For example, some models assume agents only wish to delegate to someone strictly more competent than themselves regarding the issue at hand, making directed cycles impossible.
    \item \emph{Discard cycles (DC):} Directed cycles in the graph are entirely discarded (or ignored). This is a natural solution for mechanisms using one neighbor preferences since a directed cycle in the resulting preference graph will not have any outgoing edges with paths to voters.
    \item \emph{Break cycles (BC):} The mechanism has an algorithm for breaking a directed cycle in the preference graph while preserving some non-empty subset of the edges in the cycle. For example, we could randomly choose one of the agents in Figure~\ref{fig:triangle} to vote directly while the others are allowed to delegate.
\end{itemize}

\subsubsection{Agents' right to delegate}
\label{sec:righttodelegate}

A central component of liquid democracy is that all agents are allowed to delegate their vote(s) rather than voting directly. The choice to either vote directly or delegate is essential to the motivation that it serves as a compromise between direct and representative democracy. Further, protecting an agent's right to delegate reduces the burden of requiring backup voting preferences from agents who prefer to delegate or the risk of disenfranchising agents whose delegation preferences are not honored. Regardless, it may be impossible to satisfy every agent's desire to delegate (and have their delegated vote eventually cast) due to issues such as cycles in the preference graph as discussed in Section~\ref{sec:cycles}. We can see in Figure~\ref{fig:triangle} that at least one of the three agents must forfeit this right.

However, in defining a formal ``right to delegate'', we can sidestep the unavoidable issue of cycles in worst-case analysis by focusing on the agents whose preferences combined with the preferences of others will allow for some delegation that ultimately results in votes being cast. We say that an agent has \emph{delegable preferences} if there is a directed path in the preference graph from that agent to some voter. Thus, an agent with delegable preferences has at least one simple delegation path and all such agents could simultaneously be allowed to delegate according to their preferences. 
This leads to the following property which can apply to unranked or ranked preferences.
\begin{itemize}
    \item \emph{Right to delegate (RTD):} A mechanism satisfies the \emph{right to delegate} property if every agent with delegable preferences has all of the votes assigned to them delegated to one or more of their approved delegates.
\end{itemize}

\subsubsection{Agents' right to top rank}
\label{sec:toprank}

When agents specify a ranking, we may also wish to guarantee that a mechanism respects those ranks to some reasonable extent. In theory, a mechanism could completely ignore ranks, although we are not aware of such an audacious approach. More commonly, a mechanism may make some sacrifice in respecting ranks in the service of other constraints or objectives. For example, if an agent's top-ranked delegate abstains or is part of a directed cycle, using a lower-ranked delegate would be preferable to not delegating/voting at all (See Figure~\ref{fig:GV} in Section~\ref{sec:GV} for a detailed illustration). Keeping this in mind, we establish a \emph{right to top rank} property which essentially guarantees that agents can delegate along a path using only top-ranked edges if such a path exists after removing some unusable edges from the graph.

We start by assuming some minor preprocessing of the preference graph to remove all agents which have no path to a voter. Let the \emph{delegable preference graph} be a graph constructed by removing all agents without delegable preferences from the preference graph\footnote{An algorithm is not required to explicitly construct a delegable preference graph to satisfy right to top rank, although such a graph is easy to construct.}. An agent has a \emph{top rank delegation path} in the delegable preference graph if there is a path from the agent to a voter in the delegable preference graph using only the top-ranked outgoing edges of each agent. Note that this path may use edges which were not top-ranked outgoing edges in the original preference graph and that this path is guaranteed to be simple. 
This yields a natural property for ranked preferences.
\begin{itemize}
    \item \emph{Right to top rank (RTTR):} A mechanism satisfies the \emph{right to top rank} property if every agent with a top rank delegation path has all of the votes assigned to them delegated along that path.
\end{itemize}

We note that the right to top rank and right to delegate properties do not imply each other.

\subsubsection{Limiting concentrated power}
\label{sec:concentratedpower}

Analysis of liquid democracy in practice has revealed that some individual agents will accumulate a large number of votes, becoming \emph{super-voters}\footnote{The term \emph{super-voter} is used in prior work to denote any agent accumulating a large number of votes, regardless of whether they choose to cast those votes directly or delegate.}~\citep{kling2015liquidfeedback}. Several works have argued that these super-voters are or could be detrimental. Within the theoretical model of voter competence in \citet{kahng2021algorithmic}, super-voters can make liquid democracy less likely than direct democracy to choose the correct outcome in the worst-case due to a wisdom of the crowd effect. On the practical side, \citet{kling2015liquidfeedback} observed that the presence of super-voters may have discouraged other voters from participating, but ultimately concluded that, ``While we find that the theoretical and potential power of super-voters is indeed high, we also observe that they stabilise the voting system and prevent stagnation while they use their power wisely. Super-voters do not fully act on their power to change the outcome of votes, and they vote in favour of proposals with the majority of voters in many cases.''

Concerns regarding super-voters have led to the creation of mechanisms that restrict or minimize vote accumulation~\citep{kahng2021algorithmic,golz2018fluid}. We do not take a position on the merits of limiting concentrated power, but rather explore how approaches to this goal interact with other properties we study.

For our purpose, we define two broad categories of approaches to limiting power:
\begin{itemize}
    \item \emph{Capped power (CP):} A mechanism that caps power will place a hard upper bound on the weight or number of votes than can be delegated to a single individual.
    \item \emph{Minimized power (MP):} A mechanism that minimizes power will seek to minimize the weight or number of votes than are delegated to a single individual while respecting all constraints imposed by the mechanism. 
\end{itemize}
We will see in Section~\ref{sec:fluid} that minimizing power can lead to an opaque system with some deficiencies in transparency and accountability. On the other hand, capping power can limit the ability of agents to delegate when they prefer to do so and is incompatible with guaranteeing the right to delegate. In other words, a mechanism cannot have both capped power and a right to delegate. 

\begin{theorem}
No mechanism using one neighbor preferences, multiple ranked prefences, or multiple unranked preferences can satisfy both the right to delegate and capped power for any cap less than the number of agents $n$.
\end{theorem}
\begin{proof}
This can easily be seen with just two agents. A cap of less than two will not allow either of them to delegate to the other.
\end{proof}

\subsubsection{Avoiding arbitrary decisions by the mechanism}
\label{sec:arbitrary}

Situations can arise in which the mechanism must choose arbitrarily between multiple valid delegations that would result in different outcomes for the vote or different subsets of agents being allowed to delegate. In other words, there can be a preference graph input such that a given mechanism must make a completely arbitrary decision which decides the outcome of the vote or decides who can or cannot delegate. This occurs most naturally in mechanisms that allow multiple unranked preferences, but can occur with other types of delegation preferences as well.

We view arbitrary decision-making as a negative characteristic. As such, we frame three desirable properties in this area followed by a catch-all less desirable property:
\begin{itemize}
    \item \emph{Single delegation (SD):} A mechanism is a single delegation mechanism if every set of agent preferences has exactly one allowable delegation which satisfies the constraints and objective of the mechanism. I.e., there is a many-to-one mapping from preference graphs to delegations.
    \item \emph{Single distribution over delegations (SDOD):} A randomized mechanism satisfies SDOD if every set of agent preferences has exactly one allowable distribution over delegations which satisfies the constraints and objective of the mechanism. I.e., there is a many-to-one mapping from preference graphs to distributions on delegations.
    \item \emph{No arbitrary decisions (NAD):} If there exists a set of agent preferences such that 1) the constraints and objective of the mechanism allow for more than one valid delegation (or more than one valid distribution over delegations for a randomized algorithm); 2) the mechanism deterministically chooses a delegation (or distribution); and 3) the choice of delegations affects the outcome, then we say the mechanism makes an \emph{arbitrary decision}. A mechanism satisfying the NAD property must not make an arbitrary decision, possibly by satisfying one of the previous two properties.
    \item \emph{Arbitrary:} We call a mechanism arbitrary if there exists a preference graph input on which it makes an arbitrary decision as defined in the previous property.
\end{itemize}

Clearly, single delegation is a special case of single distribution over delegations. Both satisfy the no arbitrary decisions property assuming that the problem of finding the delegation or distribution over delegations for a given preference graph input is tractable.

Note that these properties are defined in terms of the objective and constraints of the mechanism rather than the tally algorithm. In other words, a given tally algorithm may create a many-to-one mapping from preference graphs to delegations (e.g., due to breaking ties based on some arbitrary ordering of the agents), but this is not enough to satisfy the single delegation property.

\subsection{Transparency, explainability, and accountability}
\label{sec:transexplainaccount}

In the context of liquid democracy, we explore transparency, explainability, and accountability by asking how well agents can understand what has happened with their vote(s), why it happened, and what will happen if they make specific choices in the future. This information allows agents to make informed decisions about their preferences.

\subsubsection{Post hoc explainability} 
\label{sec:posthocexplain}

Properties of post hoc explainability capture the ideal in liquid democracy that agents are able to understand the outcome of their preference choices in order to inform their future preferences and delegation. Knowing where their vote(s) went is one piece of information that an agent needs to refine their preferences between voting rounds and allows them to hold their delegates accountable for past votes.

We consider two natural ways in which a mechanism can give agents feedback on what happened to their vote:
\begin{itemize}
    \item \emph{$\psi$-path explainability ($\psi$-PE):} We define a \emph{vote path} to be a simple path in a preference graph from an agent to an issue with a label specifying how the final agent in the path voted. A mechanism satisfies $\psi$-PE if it can provide every agent with a list of at most $\psi$ total vote paths and the fraction of that agent's vote which was delegated to each path. We will focus exclusively $1$-PE here, but use this more general definition in case it is relevant for future work. We are not aware of any mechanism satisfying $\psi$-path explainability for a bounded $\psi > 1$.
    \item \emph{Golden rule explainability (GRE):} The ``Golden Rule'' as described in \citet{hardt2014googletalk} and \citet{hardt2015google} for the Google Votes system is that an agent should be given a single path explanation for where their own vote went. However, as we show in Section~\ref{sec:GV}, the votes delegated to that agent may travel along a different path, violating the $1$-path explainability property described above.
    \item \emph{Local feedback explainability (LFE):} A mechanism satisfies LFE if it can provide every agent with the fraction of that agent's vote which was delegated to each neighbor and what percent of the vote delegated to each neighbor supported each outcome. 
\end{itemize}

We note that any mechanism satisfying $1$-path explainability will satisfy Golden Rule explainability and any mechanism satisfying Golden Rule explainability will satisfy local feedback explainability, but the reverse is not true.

\subsubsection{Transparency and accountability} 
\label{sec:localpredictable}

A crucial pillar of liquid democracy is accountability which is enforced partly by the ability of an agent to change their delegation preferences at any time. At the same time, it is important to consider transparency; i.e., what information an agent needs or can use to make this decision and how well they can anticipate the results of a change in their delegation preferences. 
This description from German news website, Der Spiegel, of the fluctuating support for super-voter Martin Haase within the German Pirate Party illustrates how agents engage with transparency and accountability in practice~\citep{becker2012derspiegel}. 
\begin{quote}
    Many pirates gave [Haase] their votes in 2010, when the party was seeking to define its position on the concept of an unconditional basic income guarantee. They were confident that Haase would support it. To their surprise, however, he transferred his votes to another party member, who voted against the motion, and was defeated.

    As a linguist, says Haase, he was also opposed to the initiative because of linguistic weaknesses. But this didn't convince his supporters, and about 50 Pirates promptly withdrew the votes they had assigned to him. Haase eventually approved a revised motion, and since then the number of members supporting him has gone up again.
\end{quote}

Often, the main factor conflicting with transparency and accountability in liquid democracy is the competing goal of privacy. An agent may wish to see the voting/delegation history of another agent before choosing them as a possible delegate, but this history may be hidden to respect privacy. While interesting, this tension between transparency and privacy tends to be independent of the mechanism choice and therefore outside the scope of this paper.

Along those lines, we would prefer to identify a transparency issue inherent in a given mechanism. To do so, we consider a model in which an agent knows the voting/delegation history of other agents, but does not know or cannot run the tally algorithm. In this model, an agent chooses potential delegates based on their history and should be able to assume that their future voting/delegation will be similar. 

More formally, we setup the following scenario. Suppose there are two rounds of votes in sequence. After the first vote, let each agent have a rating which determines what percentage of their vote went to each of the two outcomes. Let exactly one agent $a_c$ be the only agent to change their delegation or voting preferences between the first and second votes. Let $P_1$ and $P_2$ be the preferences of $a_c$ for the first and second votes, respectively. We say that $P_2$ \emph{strictly favors} an outcome $\mathcal{O}$ over $P_1$ if every agent in $P_2$ had a higher expected percentage of their vote(s) cast for $\mathcal{O}$ during the first vote than every agent in $P_1$. 
Using this scenario, we establish the following definition of \emph{local predictability} which captures the ability of an agent to predict the effect of changing their delegation preference. We note that for deterministic tally algorithms, local feedback explainability suffices to provide the expected percentage of an agent's vote(s) cast for $\mathcal{O}$ while a randomized algorithm would need to give the ex ante expected percentages in the preference graphs of previous votes. 
\begin{itemize}
    \item \emph{Locally predictable:} 
    We say that a mechanism is locally predictable if this change in the preferences of $a_c$ cannot reduce the expected vote share assigned to outcome $\mathcal{O}$ in the second vote of the scenario. The expectation is taken over any randomness in the tally algorithm.
\end{itemize}

It may seem surprising that mechanisms could violate the local predictability, and one could ask if it is too low of a bar to be useful. However, we show in Section~\ref{sec:fluid} that a reasonable proposed mechanism does not satisfy this property.

\subsection{Running time and tractability} 
\label{sec:runningtime}

Another important consideration is the running time of computing the tally. For different mechanisms, the tractability of computing the tally can range from solving an NP-hard problem to running a simple linear time algorithm. 
In addition to the usual concerns of efficiency, tractability of the tally problem can indirectly affect other properties. If a heuristic or approximation algorithm is required to tally votes and resolve delegations, then the choice of heuristic/algorithm or the larger space of acceptable solutions can potentially influence the outcome and/or reduce transparency. For example, two different approximation algorithms with the same approximation guarantees may lead to different delegations with different election outcomes, making the choice of which algorithm to use an arbitrary decision affecting the outcome.

\subsection{Properties contributed by this work}
\label{sec:novelproperties}

Many of these properties are explicitly or implicitly defined in prior work. However, they have not been collected in this way or compared through the lenses of accountability, transparency, explainability, fairness, and user agency. 
In addition, right to delegate (Section~\ref{sec:righttodelegate}), right to top rank (Section~\ref{sec:toprank}), avoiding arbitrariness of decisions (Section~\ref{sec:arbitrary}), and local predictability (Section~\ref{sec:localpredictable}) present novel properties which we are not aware of in any prior work.

\section{Analysis of existing mechanisms}
\label{sec:analysis}

Here, we analyze five mechanisms from prior work using the properties defined in Section~\ref{sec:properties} with several goals in mind. First, we aim to illustrate and justify the properties through example. Second, although this is far from an exhaustive list of prior work, it gives a representative snapshot of which combinations of properties have been achieved, highlighting future directions. Third, the analysis suggests additional impossibility results which are provided throughout.

\begin{table*}
  {\begin{tabular}{l>{\centering}p{40pt}>{\centering}p{30pt}>{\centering}p{35pt}>{\centering}p{30pt}>{\centering}p{40pt}>{\centering}p{40pt}>{\centering}p{40pt}>{\centering}p{40pt}p{40pt}<{\centering}}
    \multicolumn{10}{c}{{\huge Properties of Previously Proposed Mechanisms }} \\
    \multicolumn{1}{c}{} \\
\textbf{\LARGE Mechanism} 	& {\small\textbf{Prefere-\break nces}} 		& {\small\textbf{{Limit\break Power}}} 	& {\small\textbf{Outcomes}}  	& {\small\textbf{Cycles}}  	& {\small\textbf{{Right to\break Delegate}}}  	& {\small\textbf{{Right to\break Top Rank}}}  	& {\small\textbf{{Explain-\break ability}}} 	& {\small\textbf{{Locally\break Predictable}}}  	& {\small\textbf{{Running\break Time}}}  \\ 
\midrule
\textbf{Google votes} 		& {MRP} 	& {No} 		& {SD}		& {BC}		& {Yes}		& {Yes**}		& {Golden\break Rule}		& {Unclear}		& {$\Omega(2^n)$*} \\ 
\hline
\textbf{LiquidFeedback} & {ONP} 	& {No} 		& {SD}		& {DC}		& {Yes}		& {N/A}		& {1-PE}		& {Yes}		& {$O(n)$} \\ 
\hline
\textbf{Breadth-first} & {MRP} 	& {No} 		& {SD}		& {BC}		& {Yes}		& {No}		& {1-PE}		& {Yes}		& {$O(n + m)$*} \\ 
\hline
\textbf{GreedyCap}  & {MUP} 	& {CP} 		& {SDOD**}		& {AAC}		& {No}		& {N/A}		& {$1$-PE}		& {Yes}		& {$O(n \lg{n})$*}  \\
\hline
\textbf{Fluid mechanics}  & {MUP} 	& {MP} 		& {Arbitrary}		& {BC}		& {Yes}		& {N/A}		& {1-PE}		& {No}		& {Varies}  \\
    \bottomrule
    \multicolumn{1}{l}{} \\
   \end{tabular}}
   
 \caption{Summary of properties for five prior mechanisms analyzed in this section. * indicates a running time not stated in prior work, but easily derived. The Google votes running time is a lower bound on the worst case performance. ** indicates a property which can be achieved with a minor extension or resolution to an ambiguous step described in the section covering that algorithm.}\label{tab:overview}

 \end{table*}

\subsection{Google Votes}
\label{sec:GV}

The Google Votes mechanism, mentioned in \citet{hardt2015google} and described in more detail in a Google TechTalk~\citep{hardt2014googletalk}, is a liquid democracy system that was used internally within Google for Google employees to vote on issues such as meals, snacks, and project logos. It was implemented as a web app on the internal corporate Google+ network, and was in use from March 2012 to February 2015. In practice, Google Votes allowed users to express preferences for multiple issues and categories of issues all at once. However, we focus on the special case of a single binary issue for simplicity and Google Votes clearly matches the multiple ranked preferences property in this case. It is also easy to see that Google Vote does not attempt to limit concentrations of power.

The Google Votes paper~\citep{hardt2015google} defines the ``Golden Rule of Liquid Democracy'' as one of the key properties of the Google Votes system. Similar to 1-path explainability, this property states that an agent who delegates their vote must be able to see the entire path that their vote travels on, including the final vote it was cast for. The Google Votes tally algorithm achieves this by selecting one final vote path per agent. However, there are some ambiguities in the algorithm definition which make it difficult to determine whether Google Votes is actually guaranteed to satisfy 1-path explainability. It appears that their ``Golden Rule of Liquid Democracy'' applies only to the single vote that an agent starts with, rather than all votes delegated to an agent.

Google Votes resolves preferences with a tally algorithm which first enumerates all simple paths from from every agent to some voter. As such, the worst case running time is lower bounded by the number of paths which is exponential in the number of agents $n$. Then, for each agent $a$, it identifies that agent's highest ranking neighbor $a'$ for which there exists at least one delegation path starting with $a'$. From this point in the algorithm, it is not possible to determine from existing documentation how Google Votes chooses among multiple potential paths. Nevertheless, we will analyze properties based on two potential implementations of the ambiguous step and show an impossibility result that holds for any possible resolution of this step. We further note that for several possible resolutions of the ambiguous step, a fairly straightforward polynomial-time algorithm can be devised, but leave that outside the scope of this work.

\begin{figure}
\includegraphics{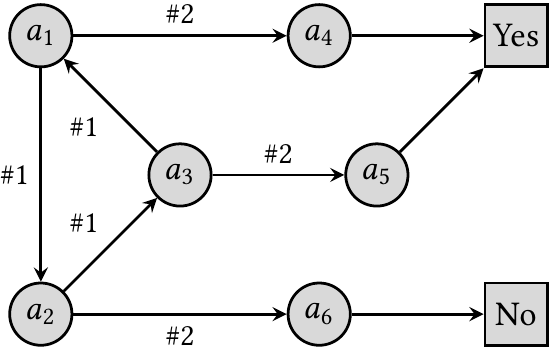}
    \caption{A preference graph that illustrates ambiguity in the the Google Votes tally algorithm.}
    \label{fig:GV}
    \Description[Graph on six agents. Three agents form a directed cycle of rank 1 edges. Each of them has a rank 2 edge to a distinct agent who is a voter]{A graph on six agents. There is a directed cycle of rank 1 edges from a1 to a2 to a3 and back. There are rank 2 edges from a1 to a4, a2 to a6, and a3 to a5. Agents a4 and a5 vote yes while a6 votes no.}
\end{figure}

To explain the issues with specifying additional properties for Google Votes, we refer to Figure~\ref{fig:GV}. This example shows three agents ($a_1$, $a_2$, and $a_3$) who all prefer to delegate to each other in a directed cycle while also having second choices ($a_4$, $a_6$, and $a_5$, respectively) who are voters. According to the limited description in \citet{hardt2014googletalk}, $a_1$ will either be given the path $(a_1,a_2,a_6)$ or $(a_1,a_2,a_3,a_5)$ based on two natural approaches that yield different delegations. Approach~1 would be to pick the path with highest ranked prefix (or highest lexicographical ordering). 
In other words, first choose the highest-ranked first step, then the highest-ranked second step, then the highest-ranked third step, and so on. 
This approach would choose the path $(a_1,a_2,a_3,a_5)$ for $a_1$ since $a_2$ prefers $a_3$ over $a_6$, and it clearly satisfies right to top rank.. Approach~2 would be to choose the shortest path from an agent which starts with its highest ranked neighbor who has some valid path that does not include the agent itself (breaking ties in length by highest ranked prefix). By contrast, this approach would choose the shorter path $(a_1,a_2,a_6)$ for $a_1$, and it does not satisfy right to top rank since the shortest path is not guaranteed to choose top-ranked edges. Approach~1 appears to be the most likely according to \citet{hardt2014googletalk}. However, neither approach nor any possible resolution to this ambiguous step can satisfy 1-path explainability. We formalize this with a theorem and brief proof sketch.

\begin{theorem}
Any tally algorithm which must delegate each agent's vote to its highest ranked neighbor through which a delegation path exists, only allows simple paths, and guarantees that the vote will eventually reach a voter, cannot satisfy 1-path explainability.
\end{theorem}
\begin{proof}
Consider the example in Figure~\ref{fig:GV}. Each of $a_1$, $a_2$, and $a_3$ will delegate their own vote to their top-ranked neighbor, and that vote can never return to them due to the simple path requirement. However, for votes to be cast, at least one of them must delegate some votes they receive through delegation to one of the voters ($a_4$, $a_5$, or $a_6$). Thus, at least one agent will send votes on more than one path, violating 1-path explainability.
\end{proof}

We note that the ambiguous tally algorithm of Google Votes is also discussed in \citet{kotsialou2018incentivising} which considers a similar ranked preference model with a focus on how participation is incentivised (or not). They characterize Approach 1 above as a \emph{depth-first delegation} rule and use an example similar to our Figure~\ref{fig:GV} to illustrate how an agent adding delegation preferences can alter the weights assigned to voters. They propose an alternative \emph{breadth-first delegation} rule which is almost identical to Approach 2, except that it prioritizes the shortest path starting from the agent rather than from their highest-ranked neighbor. Applying the breadth-first delegation rule~\citep{kotsialou2018incentivising} to Figure~\ref{fig:GV} would simply give the path $(a_1,a_4)$ to $a_1$. See Section~\ref{sec:bfd} for further details.

\subsection{LiquidFeedback}

LiquidFeedback~\citep{behrens2014principles} is possibly the most prominent real world implementation of liquid democracy and also one of the most simple approaches in terms of its mechanism. It has been used in practice (most notably by the German Pirate Party) and is available as open-source software\footnote{https://liquidfeedback.com/en/} which includes many features beyond simply implementing liquid democracy. A hallmark of LiquidFeedback's take on liquid democracy is a light touch approach that favors minimal influence of a centralized mechanism. Agents specify one neighbor preferences and delegation cycles are simply ignored by the tally algorithm (agents are expected to resolve cycles themselves by at least one member of the cycle becoming a voter). Similarly, it satisfies the right to delegate and there is no attempt to limit concentration of power. After cycles are removed, the preference graph becomes a directed tree with a maximum out-degree of one. This easily implies 1-path explainability, single delegation, and no arbitrary decisions.

\subsection{Breadth-first delegation}
\label{sec:bfd}

As noted in Section~\ref{sec:GV}, \citet{kotsialou2018incentivising} proposed \emph{breadth-first delegation} as an alternative to ``depth-first'' approaches such as Google Votes for tallying multiple ranked preferences. A breadth-first delegation chooses the shortest path from an agent to some voter and only uses edge ranks to break ties amongst multiple paths of the same length. 
In motivating this approach, \citet{kotsialou2018incentivising} argue, for example, that an agent may prefer to delegate to their second choice over their top choice when their second choice is a voter, but their top choice will transitively delegate. Thus, breadth-first delegation explicitly chooses not to satisfy the right to top rank. In fact, if an agent's worst-ranked delegate is a voter while all better ranked delegates are agents who delegate to voters, then a breadth-first tally will choose the worst-ranked delegate. 
We acknowledge that this preference may exist in some settings and describe some advantages offered by a breadth-first tally, but note that it goes against the traditional model of liquid democracy and violates the explicit ranking preferences of the agent. Delegating to someone else who then further chooses whom to delegate to (an expert on selecting experts) is fundamental to liquid democracy and a major feature distinguishing it from proxy voting.

Despite this key philosophical difference of prioritizing path length over ranks, breadth-first delegation shares many properties in common with Google Votes. By choosing the unique shortest path which has the highest lexicographical ordering of ranks for each agent, it ensures that each preference graph implies a single delegation with the right to delegate and break cycles properties. It also does not limit power in the worst case. However, \citet{kotsialou2018incentivising} suggest the open question of whether the shorter paths will accumulate fewer votes on real world input and have the effect of limiting concentrated power in practice.

On the other hand, breadth-first delegation differs from Google Votes by offering $1$-path explainability and local predictability. Lemma 2 of \citet{kotsialou2018incentivising} implies $1$-path explainability by showing that if one agent is on the chosen delegation path of another agent, then both of them are delegating to the same voter. 
To briefly sketch why local predictability holds, consider an agent $a_c$ with distinct sets of preferences $P_1$ and $P_2$ for the first and second votes, respectively. Let $\mathcal{O}$ be the outcome strictly favored by $P_2$ and $a_d \in P_2$ be the agent whom $a_c$ ultimately delegates to in the second vote. By the single delegation property and 1-path explainability, $a_c$ and $a_d$ and must have had disjoint delegation paths that delegated all of their assigned votes to different outcomes in the first vote. It follows that $a_d$ will have the same delegation path in the second vote as well and will not lose any of the votes delegated to them in the first vote although some votes may now travel to $a_d$ through $a_c$. Thus, in the second vote, outcome $\mathcal{O}$ will receive all of the votes it received in the first vote plus at least one additional vote (from $a_c$).

\subsection{The GreedyCap algorithm}

The GreedyCap algorithm was introduced in \citet{kahng2021algorithmic} to prove that there exists a non-local delegation mechanism which satisfies their properties of positive gain and do no harm with respect to direct democracy. Roughly speaking, \emph{positive gains} means there is some preference graph on which a given liquid democracy mechanism outperforms direct voting at finding the ground truth in a model in which there is one correct outcome. \emph{Do no harm} is the complementary property stating that the liquid democracy mechanism does not perform worse than direct democracy.

The input to GreedyCap is an unweighted, directed graph with vertices representing agents and a cap $C$ on the number of votes that can be delegated to a single agent. A directed edge $(u, v)$ signifies that agent $u$ approves of delegating their vote to agent $v$. This simple greedy algorithm proceeds by iteratively choosing the remaining agent $v$ with the most approvals (incoming edges) and at most $C-1$ agents that approve of $v$. These agents (including $v$) are removed from the graph with their votes all being delegated to $v$. At the end of this process, all agents whose votes have not been delegated to someone else are instructed to vote.

We can now take a look at which properties are satisfied by GreedyCap. It is easy to see that the input graph here satisfies the multiple unranked preference property, and power is strictly capped by the parameter $C$. Cycles are not considered because \citet{kahng2021algorithmic} assumes a model in which each agent has a competence value and will only delegate to someone strictly more competent, making this a mechanism that assumes away cycles. Explainability is not considered in \citet{kahng2021algorithmic}. However, because this mechanism only applies ``one hop'' of delegation, it is technically a form of proxy voting and clearly offers $1$-path explainability. Similarly it also satisfies local predictability since all agents are delegating directly to voters. However, we note that generalizing this approach of capping power to true liquid democracy may result in the loss of local predictability. In fact, it is easy to see that some natural mechanisms which cap voter power would not have local predictability, and we leave this as an exercise for the reader.

Implementing this algorithm as defined in \citet{kahng2021algorithmic} could make arbitrary decisions that affect the outcome. There is no explanation in that paper for how ties are broken at various stages (it isn't relevant to the theorem they're proving). However, if these ties are broken uniformly at random, we get a single distribution over delegations.

\subsection{The fluid mechanics approach}
\label{sec:fluid}

The fluid mechanics approach~\citep{golz2018fluid} takes inspiration from fluid mechanics to address the issue of concentrated power. They seek to minimize the maximum power of any voter using a flow-based tally algorithm with the following high-level idea: ``Put another (more whimsical) way, we wish to design liquid democracy systems that emulate the \emph{law of communicating vessels}, which asserts that liquid will find an equal level in connected containers''

The input is an unweighted, directed graph (multiple unranked preferences) with each agent being a \emph{source} of one vote and all voters being \emph{sinks}. The tally algorithm searches for a confluent flow in this graph to establish the delegation. In addition to the common \emph{flow conservation} property, a confluent flow has the \emph{confluence} property that at most one outgoing edge from each vertex may carry nonzero flow. Finding such a flow is NP-hard leading to three potential tally options described in \citet{golz2018fluid}: finding an exact solution using a MILP on sufficiently small instances, a heuristic, or an approximation algorithm. As such, the running time varies greatly depending on specific implementation choices.

The reasoning for tackling the NP-hard confluent flow problem rather than an easier splittable flow variant is to allow for 1-path explainability. Although \citet{golz2018fluid} does not explicitly discuss giving agents single-path explanations, they note that confluent flow is more intuitive and transparent leading to higher accountability. Further, their experimental results do not show a significant decrease in maximum power when switching to splittable flow.

In addition to 1-path explainability, it is easy to see that this method provides a right to delegate. Any agent with a path to some voters, will send flow to one of those voters. Similarly, cycles are broken if at least one agent in the cycle has some other delegation path, while agents connected components with no path to any voter will be discarded (alternately, they could be allowed to vote directly, but this is not discussed in the paper).

\begin{figure}
\includegraphics{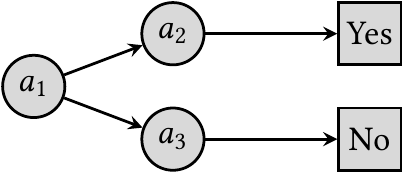}
    \caption{Illustrating a preference graph which forces the fluid mechanics approach to arbitrarily choose the outcome of the vote. The vote of $a_1$ must be delegated to either $a_2$ or $a_3$. Both options are optimal with respect to the objective and confluence constraint and this arbitrary choice determines whether the outcome will be ``yes'' or ``no''.}
    \label{fig:arbitraryflow}
    \Description{Agent a1 has directed edges to a2 who votes yes and a3 who votes no.}
\end{figure}

One potential drawback to this system and using confluent flows is that some preference graphs have multiple distinct delegation graphs which optimize the minimum confluent flow objective. Figure~\ref{fig:arbitraryflow} illustrates a simple example of this. The vote of $a_1$ must be delegated to either $a_2$ or $a_3$. Both options are optimal with respect to the objective and confluence constraint and this arbitrary choice determines whether the outcome will be ``yes'' or ``no''. Thus, this method can potentially make arbitrary decisions that affect the outcome of a vote. We can formalize and generalize this observation with the following theorem which states that a deterministic mechanism cannot guarantee multiple unranked preferences, the right to delegate, single-path explainability, and no arbitrary decisions.
\begin{theorem}
\label{thm:unrankedarbitrary}
No deterministic mechanism with multiple unranked preferences can simultaneously satisfy the right to delegate, single-path explainability, and no arbitrary decisions.
\end{theorem}
\begin{proof}
The proof follows from the above discussion of Figure~\ref{fig:arbitraryflow}.
\end{proof}

\begin{figure*}
\includegraphics{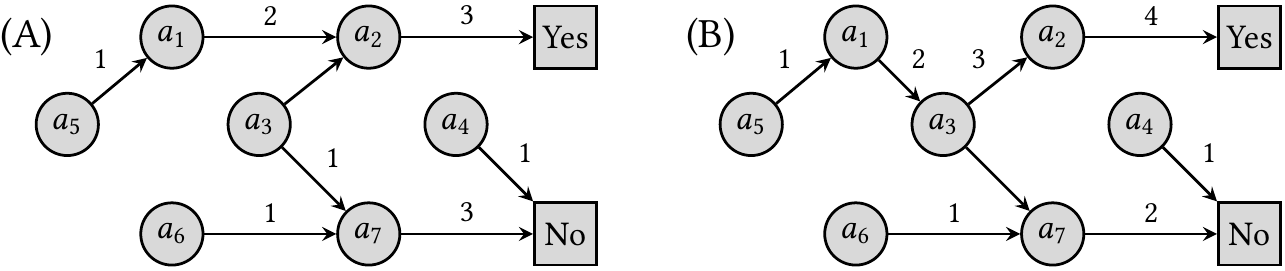}
    \caption{A pair of preference graphs showing how the Fluid Mechanics approach can produce an output contrary to an agent's intention. The numbers above each edge indicate the amount of flow traveling through the edge after minimizing confluent flow. (A) We see the first vote in which $a_1$ prefers to delegate only to $a_2$ who votes for ``yes'', but ``no'' wins. (B) We see the second vote in which $a_1$ changes their preference to $a_3$ who had voted for ``no'' in the first vote, but this change causes ``yes'' to win.}
    \label{fig:reverseflow}
    \Description{Graph (A) shows delegation paths (a5, a1, a2, yes), (a4, no), (a3, a7, no), and (a6, a7, no). The only unused edge in (A) is (a3, a2). Graph (B) shows delegation paths (a5, a1, a3, a2, yes), (a4, no), and (a6, a7, no). The only unused edge in (B) is (a3, a7).}
\end{figure*}

Another issue is the potential to produce outcomes that are counter to an agent's intention and violate the locally predictable property. The preference graphs in Figure~\ref{fig:reverseflow} and their minimum confluent flows illustrate this effect. In (A), we see the first vote in which $a_1$ prefers to delegate only to $a_2$ who votes for ``yes'', but ``no'' ultimately wins (3 votes for ``yes'' and 4 for ``no''). Now, suppose $a_1$ decides that ``no'' was truly the better choice and decides they would rather delegate to $a_3$ who's vote previously followed a path to ``no'' in the first vote (A). Then, in (B), we see a second vote in which $a_1$ has changed their preference to $a_3$, but this change causes ``yes'' to win (4 votes for ``yes'' and 3 for ``no'').

\section{Conclusion and future directions}
\label{sec:conclusion}

The analysis in Section~\ref{sec:analysis} suggests that ideals of fairness, explainability, transparency, and accountability can be at odds with goals of limiting concentrated power or selecting the ground truth correct outcome. A similar divide is present between practical implementations~\citep{hardt2015google,behrens2014principles} and theoretical work~\citep{kahng2021algorithmic,golz2018fluid} with the former giving less influence to a centralized mechanism while the latter type tend to explore how a centralized mechanism could exert a beneficial influence on the system. Given these trade-offs between desirable traits, it would be useful to better understand what other combinations of properties can or cannot be achieved.

Applications similar to those which have used Google Votes or LiquidFeedback might prefer a mechanism with certain desirable properties of bread-first delegation, but prioritizing edge ranks over shortest paths. In addition to multiple ranked preferences, SD, BC, right to delegate, right to top rank, 1-PE, local predictability, and polynomial running time, it would satisfy right to top rank and/or some other guarantee relating to edge ranks. This implies the additional open question of what other properties we could guarantee for giving agents their best ranked paths. E.g., one might incorporate an egalitarian objective which maximizes the minimum rank of any edge in the delegation graph.

When minimizing power is a high priority, we can ask if it's possible to minimize power in some way while ensuring local predictability and SD or SDOD. This might involve using multiple ranked preferences or abandoning 1-PE to satisfy SD to sidestep the impossibility result of Theorem~\ref{thm:unrankedarbitrary}. Moreover, as suggested by \citet{kotsialou2018incentivising}, an empirical study could reveal which mechanisms tend to yield lower concentrations of power on real or simulated data.

On the epistemic side, there is the question of whether a liquid democracy mechanism can provide strict improvement over direct democracy while achieving features not possessed by GreedyCap, such as a right to delegate. Due to the negative result of \citet{kahng2021algorithmic}, the most promising direction might involve developing and justifying a new model of voter competence that is not subject to the same impossibility.

Finally, the approach of this paper could be extended to other democratic systems in which computation plays a role. For example, the computer science and mathematics communities have given increasing attention to redistricting and gerrymandering in district-based representative democracies. Algorithms have been proposed to draw congressional district maps, thereby controlling the delegation of citizens’ votes to representatives and affecting election outcomes~\citep{Liu2016,cohen2018balanced,gurnee2021fairmandering,esmaeili2022centralized}. They are also used to measure gerrymandering in existing maps~\citep{mattingly2014redistricting,Chikina2017,herschlag2020quantifying,deford2019recombination} and have been shown to affect theoretical guarantees on voter incentives~\citep{brubach2020meddling}.


\bibliographystyle{ACM-Reference-Format}
\bibliography{refs}


\begin{thebibliography}{26}


\ifx \showCODEN    \undefined \def \showCODEN     #1{\unskip}     \fi
\ifx \showDOI      \undefined \def \showDOI       #1{#1}\fi
\ifx \showISBNx    \undefined \def \showISBNx     #1{\unskip}     \fi
\ifx \showISBNxiii \undefined \def \showISBNxiii  #1{\unskip}     \fi
\ifx \showISSN     \undefined \def \showISSN      #1{\unskip}     \fi
\ifx \showLCCN     \undefined \def \showLCCN      #1{\unskip}     \fi
\ifx \shownote     \undefined \def \shownote      #1{#1}          \fi
\ifx \showarticletitle \undefined \def \showarticletitle #1{#1}   \fi
\ifx \showURL      \undefined \def \showURL       {\relax}        \fi
\providecommand\bibfield[2]{#2}
\providecommand\bibinfo[2]{#2}
\providecommand\natexlab[1]{#1}
\providecommand\showeprint[2][]{arXiv:#2}

\bibitem[\protect\citeauthoryear{Becker}{Becker}{2012}]%
        {becker2012derspiegel}
\bibfield{author}{\bibinfo{person}{Sven Becker}.}
  \bibinfo{year}{2012}\natexlab{}.
\newblock \showarticletitle{``Web Platform Makes Professor Most Powerful
  Pirate''}.
\newblock \bibinfo{journal}{\emph{Der Spiegal}} (\bibinfo{year}{2012}).
\newblock
\urldef\tempurl%
\url{https://www.spiegel.de/international/germany/liquid-democracy-web-platform-makes-professor-most-powerful-pirate-a-818683.html}
\showURL{%
\tempurl}
\newblock
\shownote{[Online; accessed 24-April-2022]}.


\bibitem[\protect\citeauthoryear{Behrens, Kistner, Nitsche, and
  Swierczek}{Behrens et~al\mbox{.}}{2014}]%
        {behrens2014principles}
\bibfield{author}{\bibinfo{person}{Jan Behrens}, \bibinfo{person}{Axel
  Kistner}, \bibinfo{person}{Andreas Nitsche}, {and} \bibinfo{person}{Bj{\"o}rn
  Swierczek}.} \bibinfo{year}{2014}\natexlab{}.
\newblock \bibinfo{booktitle}{\emph{The principles of LiquidFeedback}}.
\newblock \bibinfo{publisher}{Interaktive Demokratie e. V. Berlin}.
\newblock


\bibitem[\protect\citeauthoryear{Blum and Zuber}{Blum and Zuber}{2016}]%
        {blum2016liquid}
\bibfield{author}{\bibinfo{person}{Christian Blum} {and}
  \bibinfo{person}{Christina~Isabel Zuber}.} \bibinfo{year}{2016}\natexlab{}.
\newblock \showarticletitle{Liquid democracy: Potentials, problems, and
  perspectives}.
\newblock \bibinfo{journal}{\emph{Journal of Political Philosophy}}
  \bibinfo{volume}{24}, \bibinfo{number}{2} (\bibinfo{year}{2016}),
  \bibinfo{pages}{162--182}.
\newblock


\bibitem[\protect\citeauthoryear{Brill}{Brill}{2018}]%
        {brill2018interactive}
\bibfield{author}{\bibinfo{person}{Markus Brill}.}
  \bibinfo{year}{2018}\natexlab{}.
\newblock \showarticletitle{Interactive democracy}. In
  \bibinfo{booktitle}{\emph{Proceedings of the 17th International Conference on
  Autonomous Agents and MultiAgent Systems}}. \bibinfo{pages}{1183--1187}.
\newblock


\bibitem[\protect\citeauthoryear{Brill and Talmon}{Brill and Talmon}{2018}]%
        {brill2018pairwise}
\bibfield{author}{\bibinfo{person}{Markus Brill} {and} \bibinfo{person}{Nimrod
  Talmon}.} \bibinfo{year}{2018}\natexlab{}.
\newblock \showarticletitle{Pairwise Liquid Democracy.}. In
  \bibinfo{booktitle}{\emph{IJCAI}}, Vol.~\bibinfo{volume}{18}.
  \bibinfo{pages}{137--143}.
\newblock


\bibitem[\protect\citeauthoryear{Brubach, Srinivasan, and Zhao}{Brubach
  et~al\mbox{.}}{2020}]%
        {brubach2020meddling}
\bibfield{author}{\bibinfo{person}{Brian Brubach}, \bibinfo{person}{Aravind
  Srinivasan}, {and} \bibinfo{person}{Shawn Zhao}.}
  \bibinfo{year}{2020}\natexlab{}.
\newblock \showarticletitle{Meddling metrics: the effects of measuring and
  constraining partisan gerrymandering on voter incentives}. In
  \bibinfo{booktitle}{\emph{Proceedings of the 21st ACM Conference on Economics
  and Computation}}. \bibinfo{pages}{815--833}.
\newblock


\bibitem[\protect\citeauthoryear{Caragiannis and Micha}{Caragiannis and
  Micha}{2019}]%
        {caragiannis2019contribution}
\bibfield{author}{\bibinfo{person}{Ioannis Caragiannis} {and}
  \bibinfo{person}{Evi Micha}.} \bibinfo{year}{2019}\natexlab{}.
\newblock \showarticletitle{A Contribution to the Critique of Liquid
  Democracy.}. In \bibinfo{booktitle}{\emph{IJCAI}}. \bibinfo{pages}{116--122}.
\newblock


\bibitem[\protect\citeauthoryear{Carroll}{Carroll}{1884}]%
        {carroll1884principles}
\bibfield{author}{\bibinfo{person}{Lewis Carroll}.}
  \bibinfo{year}{1884}\natexlab{}.
\newblock \bibinfo{booktitle}{\emph{The principles of parliamentary
  representation}}.
\newblock \bibinfo{publisher}{Harrison and Sons}.
\newblock


\bibitem[\protect\citeauthoryear{Chikina, Frieze, and Pegden}{Chikina
  et~al\mbox{.}}{2017}]%
        {Chikina2017}
\bibfield{author}{\bibinfo{person}{Maria Chikina}, \bibinfo{person}{Alan
  Frieze}, {and} \bibinfo{person}{Wesley Pegden}.}
  \bibinfo{year}{2017}\natexlab{}.
\newblock \showarticletitle{Assessing significance in a Markov chain without
  mixing}.
\newblock \bibinfo{journal}{\emph{Proceedings of the National Academy of
  Sciences}} \bibinfo{volume}{114}, \bibinfo{number}{11}
  (\bibinfo{year}{2017}), \bibinfo{pages}{2860--2864}.
\newblock
\showISSN{0027-8424}


\bibitem[\protect\citeauthoryear{Cohen-Addad, Klein, and Young}{Cohen-Addad
  et~al\mbox{.}}{2018}]%
        {cohen2018balanced}
\bibfield{author}{\bibinfo{person}{Vincent Cohen-Addad},
  \bibinfo{person}{Philip~N Klein}, {and} \bibinfo{person}{Neal~E Young}.}
  \bibinfo{year}{2018}\natexlab{}.
\newblock \showarticletitle{Balanced centroidal power diagrams for
  redistricting}. In \bibinfo{booktitle}{\emph{Proceedings of the 26th ACM
  SIGSPATIAL International Conference on Advances in Geographic Information
  Systems}}. \bibinfo{pages}{389--396}.
\newblock


\bibitem[\protect\citeauthoryear{Colley, Grandi, and Novaro}{Colley
  et~al\mbox{.}}{2021}]%
        {colley2021smart}
\bibfield{author}{\bibinfo{person}{Rachael Colley}, \bibinfo{person}{Umberto
  Grandi}, {and} \bibinfo{person}{Arianna Novaro}.}
  \bibinfo{year}{2021}\natexlab{}.
\newblock \showarticletitle{Smart voting}. In
  \bibinfo{booktitle}{\emph{Twenty-Ninth International Joint Conference on
  Artificial Intelligence (IJCAI 2020)}}. International Joint Conferences on
  Artifical Intelligence (IJCAI), \bibinfo{pages}{1734--1740}.
\newblock


\bibitem[\protect\citeauthoryear{DeFord, Duchin, and Solomon}{DeFord
  et~al\mbox{.}}{2019}]%
        {deford2019recombination}
\bibfield{author}{\bibinfo{person}{Daryl DeFord}, \bibinfo{person}{Moon
  Duchin}, {and} \bibinfo{person}{Justin Solomon}.}
  \bibinfo{year}{2019}\natexlab{}.
\newblock \showarticletitle{Recombination: A family of Markov chains for
  redistricting}.
\newblock \bibinfo{journal}{\emph{arXiv preprint arXiv:1911.05725}}
  (\bibinfo{year}{2019}).
\newblock


\bibitem[\protect\citeauthoryear{Esmaeili, Grape, and Brubach}{Esmaeili
  et~al\mbox{.}}{2022}]%
        {esmaeili2022centralized}
\bibfield{author}{\bibinfo{person}{Seyed~A Esmaeili}, \bibinfo{person}{Hayley
  Grape}, {and} \bibinfo{person}{Brian Brubach}.}
  \bibinfo{year}{2022}\natexlab{}.
\newblock \showarticletitle{Centralized Fairness for Redistricting}.
\newblock \bibinfo{journal}{\emph{arXiv preprint arXiv:2203.00872}}
  (\bibinfo{year}{2022}).
\newblock


\bibitem[\protect\citeauthoryear{G{\"o}lz, Kahng, Mackenzie, and
  Procaccia}{G{\"o}lz et~al\mbox{.}}{2018}]%
        {golz2018fluid}
\bibfield{author}{\bibinfo{person}{Paul G{\"o}lz}, \bibinfo{person}{Anson
  Kahng}, \bibinfo{person}{Simon Mackenzie}, {and} \bibinfo{person}{Ariel~D
  Procaccia}.} \bibinfo{year}{2018}\natexlab{}.
\newblock \showarticletitle{The fluid mechanics of liquid democracy}. In
  \bibinfo{booktitle}{\emph{International Conference on Web and Internet
  Economics}}. Springer, \bibinfo{pages}{188--202}.
\newblock


\bibitem[\protect\citeauthoryear{Gurnee and Shmoys}{Gurnee and Shmoys}{2021}]%
        {gurnee2021fairmandering}
\bibfield{author}{\bibinfo{person}{Wes Gurnee} {and} \bibinfo{person}{David~B
  Shmoys}.} \bibinfo{year}{2021}\natexlab{}.
\newblock \showarticletitle{Fairmandering: A column generation heuristic for
  fairness-optimized political districting}. In \bibinfo{booktitle}{\emph{SIAM
  Conference on Applied and Computational Discrete Algorithms (ACDA21)}}. SIAM,
  \bibinfo{pages}{88--99}.
\newblock


\bibitem[\protect\citeauthoryear{Hardt}{Hardt}{2014}]%
        {hardt2014googletalk}
\bibfield{author}{\bibinfo{person}{Steve Hardt}.}
  \bibinfo{year}{2014}\natexlab{}.
\newblock \bibinfo{title}{Liquid Democracy with Google Votes}.
\newblock
\newblock
\urldef\tempurl%
\url{https://www.youtube.com/watch?v=F4lkCECSBFw}
\showURL{%
\tempurl}


\bibitem[\protect\citeauthoryear{Hardt and Lopes}{Hardt and Lopes}{2015}]%
        {hardt2015google}
\bibfield{author}{\bibinfo{person}{Steve Hardt} {and} \bibinfo{person}{Lia~CR
  Lopes}.} \bibinfo{year}{2015}\natexlab{}.
\newblock \showarticletitle{Google votes: A liquid democracy experiment on a
  corporate social network}.
\newblock  (\bibinfo{year}{2015}).
\newblock


\bibitem[\protect\citeauthoryear{Herschlag, Kang, Luo, Graves, Bangia, Ravier,
  and Mattingly}{Herschlag et~al\mbox{.}}{2020}]%
        {herschlag2020quantifying}
\bibfield{author}{\bibinfo{person}{Gregory Herschlag},
  \bibinfo{person}{Han~Sung Kang}, \bibinfo{person}{Justin Luo},
  \bibinfo{person}{Christy~Vaughn Graves}, \bibinfo{person}{Sachet Bangia},
  \bibinfo{person}{Robert Ravier}, {and} \bibinfo{person}{Jonathan~C
  Mattingly}.} \bibinfo{year}{2020}\natexlab{}.
\newblock \showarticletitle{Quantifying gerrymandering in north carolina}.
\newblock \bibinfo{journal}{\emph{Statistics and Public Policy}}
  \bibinfo{volume}{7}, \bibinfo{number}{1} (\bibinfo{year}{2020}),
  \bibinfo{pages}{30--38}.
\newblock


\bibitem[\protect\citeauthoryear{Kahng, Mackenzie, and Procaccia}{Kahng
  et~al\mbox{.}}{2021}]%
        {kahng2021algorithmic}
\bibfield{author}{\bibinfo{person}{Anson Kahng}, \bibinfo{person}{Simon
  Mackenzie}, {and} \bibinfo{person}{Ariel Procaccia}.}
  \bibinfo{year}{2021}\natexlab{}.
\newblock \showarticletitle{Liquid democracy: An algorithmic perspective}.
\newblock \bibinfo{journal}{\emph{Journal of Artificial Intelligence Research}}
   \bibinfo{volume}{70} (\bibinfo{year}{2021}), \bibinfo{pages}{1223--1252}.
\newblock


\bibitem[\protect\citeauthoryear{Kling, Kunegis, Hartmann, Strohmaier, and
  Staab}{Kling et~al\mbox{.}}{2015}]%
        {kling2015liquidfeedback}
\bibfield{author}{\bibinfo{person}{Christoph~Carl Kling},
  \bibinfo{person}{J{\'e}r{\^o}me Kunegis}, \bibinfo{person}{Heinrich
  Hartmann}, \bibinfo{person}{Markus Strohmaier}, {and}
  \bibinfo{person}{Steffen Staab}.} \bibinfo{year}{2015}\natexlab{}.
\newblock \showarticletitle{Voting behaviour and power in online democracy: A
  study of LiquidFeedback in Germany's Pirate Party}. In
  \bibinfo{booktitle}{\emph{Ninth International AAAI Conference on Web and
  Social Media}}.
\newblock


\bibitem[\protect\citeauthoryear{Kotsialou and Riley}{Kotsialou and
  Riley}{2018}]%
        {kotsialou2018incentivising}
\bibfield{author}{\bibinfo{person}{Grammateia Kotsialou} {and}
  \bibinfo{person}{Luke Riley}.} \bibinfo{year}{2018}\natexlab{}.
\newblock \showarticletitle{Incentivising participation in liquid democracy
  with breadth-first delegation}.
\newblock \bibinfo{journal}{\emph{arXiv preprint arXiv:1811.03710}}
  (\bibinfo{year}{2018}).
\newblock


\bibitem[\protect\citeauthoryear{Liu, Cho, and Wang}{Liu et~al\mbox{.}}{2016}]%
        {Liu2016}
\bibfield{author}{\bibinfo{person}{Yan~Y. Liu}, \bibinfo{person}{Wendy K.~Tam
  Cho}, {and} \bibinfo{person}{Shaowen Wang}.} \bibinfo{year}{2016}\natexlab{}.
\newblock \showarticletitle{PEAR: a massively parallel evolutionary computation
  approach for political redistricting optimization and analysis}.
\newblock \bibinfo{journal}{\emph{Swarm and Evolutionary Computation}}
  \bibinfo{volume}{30} (\bibinfo{year}{2016}), \bibinfo{pages}{78 -- 92}.
\newblock
\showISSN{2210-6502}


\bibitem[\protect\citeauthoryear{Mattingly and Vaughn}{Mattingly and
  Vaughn}{2014}]%
        {mattingly2014redistricting}
\bibfield{author}{\bibinfo{person}{Jonathan~C Mattingly} {and}
  \bibinfo{person}{Christy Vaughn}.} \bibinfo{year}{2014}\natexlab{}.
\newblock \showarticletitle{Redistricting and the Will of the People}.
\newblock \bibinfo{journal}{\emph{arXiv preprint arXiv:1410.8796}}
  (\bibinfo{year}{2014}).
\newblock


\bibitem[\protect\citeauthoryear{Paulin}{Paulin}{2020}]%
        {paulin2020overview}
\bibfield{author}{\bibinfo{person}{Alois Paulin}.}
  \bibinfo{year}{2020}\natexlab{}.
\newblock \showarticletitle{An overview of ten years of liquid democracy
  research}. In \bibinfo{booktitle}{\emph{The 21st Annual International
  Conference on Digital Government Research}}. \bibinfo{pages}{116--121}.
\newblock


\bibitem[\protect\citeauthoryear{Ramos}{Ramos}{2015}]%
        {ramos2015liquid}
\bibfield{author}{\bibinfo{person}{Jos{\'e} Ramos}.}
  \bibinfo{year}{2015}\natexlab{}.
\newblock \showarticletitle{Liquid democracy and the futures of governance}.
\newblock In \bibinfo{booktitle}{\emph{The Future Internet}}.
  \bibinfo{publisher}{Springer}, \bibinfo{pages}{173--191}.
\newblock


\bibitem[\protect\citeauthoryear{Zhang and Grossi}{Zhang and Grossi}{2021}]%
        {zhang2021tracking}
\bibfield{author}{\bibinfo{person}{Yuzhe Zhang} {and} \bibinfo{person}{Davide
  Grossi}.} \bibinfo{year}{2021}\natexlab{}.
\newblock \showarticletitle{Tracking Truth by Weighting Proxies in Liquid
  Democracy}.
\newblock \bibinfo{journal}{\emph{arXiv preprint arXiv:2103.09081}}
  (\bibinfo{year}{2021}).
\newblock


\end{thebibliography}



\end{document}